\documentclass[12pt]{article}
\usepackage{authblk}
\usepackage{amsmath}
\usepackage{amsthm}
\usepackage{graphicx}
\usepackage{amsfonts}
\usepackage{amssymb}
\usepackage{booktabs}
\usepackage{subfigure}
\usepackage{multirow}
\usepackage{blindtext}
\usepackage{tabularx}
\usepackage{float}
\usepackage[section]{placeins}
\usepackage{cite}
\usepackage[font={footnotesize,it}]{caption}

\theoremstyle{definition}
\newtheorem*{definition*}{Definition}

\title{Adiabatic Isometric Mapping Algorithm for Embedding 2-Surfaces in Euclidean 3-Space}

\author[1]{Shannon Ray}
\author[1,2]{Warner A  Miller}
\author[3]{Paul M  Alsing} 
\author[2]{\\Shing-Tung Yau}
\affil[1]{Department of Physics, Florida Atlantic University, Boca Raton FL 33431, USA}
\affil[2]{Department of Mathematics, Harvard University, Cambridge MA 02138, USA}
\affil[3]{Information Directorate, Air Force Research Laboratory, Rome, NY 13441, USA}

\begin{document}

\maketitle

\begin{abstract}

Alexandrov proved that any simplicial complex homeomorphic to a sphere with strictly non-negative Gaussian curvature at each vertex can be isometrically embedded uniquely in $\mathbb{R}^3$ as a convex polyhedron. Due to the nonconstructive nature of his proof, there have yet to be any algorithms, that we know of, that realizes the Alexandrov embedding in polynomial time.  Following his proof, we developed the adiabatic isometric mapping (AIM) algorithm.  AIM uses a guided adiabatic pull-back procedure on a given polyhedral metric to produce an embedding that approximates the unique Alexandrov polyhedron.  Tests of AIM applied to two different polyhedral metrics suggests that its run time is sub cubic with respect to the number of vertices.  Although Alexandrov's theorem specifically addresses the embedding of convex polyhedral metrics, we tested AIM on a broader class of polyhedral metrics that included regions of negative Gaussian curvature.  One test was on a surface just outside the  ergosphere of a Kerr black hole.   

\end{abstract}

\newpage

\section{Introduction}
The problem of embedding surfaces homeomorphic to $\mathcal{S}^2$ with a metric of positive Gaussian curvature into $\mathbb{R}^3$ was posed by Herman Weyl in 1916 \cite{Weyl}.  A useful review and analysis of the isometric embedding of Riemannian manifolds in Euclidean spaces can be found in \cite{HanHong}.  The first attempt to prove the existence of an embedding was given by Weyl himself.  He was not able to complete his proof, but he did make progress in outlining a solution.  Following Weyl's approach, a proof was given by H. Lewy in 1938 though his solution required the components of the metric to be infinitely differentiable~\cite{Lewy:Weyl}.  Requirements on differentiability of the metric were reduced to continuous fourth derivatives by L. Nirenberg in 1953~\cite{Nirenberg:WeylProof}.  They were further reduced to continuous third derivatives by E. Heinz in 1962~\cite{Heinz:1962}.  
In 1941, Alexandrov gave an approach that relied on proving the existence of a convex polyhedron given any convex polyhedral metric.  He showed that in the limit as the number of vertices in the polyhedral metric goes to infinity, one recovers the metric in the continuum.  This result can be found in a compendium of his work published in 2006~\cite{Alexandrov:IntrinsicConvex}.

Our interest in isometric embeddings stems from its application in general relativity.  Solutions to Einstein's equations give a four dimensional spacetime manifold.  One can slice these manifolds in such a way to get interesting two surfaces such as the event horizon or ergosphere. To visualize these surfaces it may be best to embed them in three dimensional flat Euclidean space.  Another application of isometric embeddings include computing quasi-local energy (QLE) in general relativity.  Many definitions of quasi-local energy, such as the Brown-York and Wang-Yau energies, require isometric embeddings of convex $\mathcal{S}^2$ surfaces in $\mathbb{R}^3$~\cite{BY:QLM,WangYau:PRL}.  A detailed review of the development of QLE is given in~\cite{Szabados}. 

Though the mathematics of global isometric embeddings of convex two-surfaces homeomorphic to $\mathbb{S}^2$ was studied exhaustively for over 60 years, its numerical implementation has proven formidable and has only been approached practically since the mid-1990's. In 2006, Bondarescu et. al introduced a numerical method using the components of the metric in the continuum~\cite{Bondarescu:2002}.  They expanded the embedding functions using spherical harmonics and minimized the coefficients by optimizing the squared difference in metric components.  Bondarescu et. al found that their search would get stuck in local minimum while minimizing in the space of coefficients.  To reduce their residuals, they increased the number of coefficients for each embedding function.

In 2011, M. Jasiulek and M. Korzy\'{n}ski introduced an algorithm that closely followed the continuity method prescribed by Weyl~\cite{JK:2012}.  First they uniformized their surface using Ricci flow, which gave them conformal relations between their original metric, the round sphere metric and all intermediate metrics.  They then embedded the spherical metric into $\mathbb{R}^3$.  This allowed them to use the embedded sphere as a starting point for solving the linearized embedding equations which gave them the deformation in coordinates such that they move from the sphere to a sufficiently close conformally related surface.  Because they were working in the continuum, they dealt with typical difficulties such as finding suitable coordinate charts, solving non-linear elliptical PDE's and computing integrals and derivatives of functions on the surface.  They were also restricted to only embedding surfaces with strictly positive curvature due to the inversion of the extrinsic curvature tensor in their procedure. 

Working in the continuum has the added complication of requiring prior knowledge about the sign of the Gaussian curvature of the surface.  Whether or not the embedding equations are elliptical or hyperbolic depends on the sign of the curvature.  Additionally one may need to provide numerical techniques to deal with coordinate singularities.  Difficulties in the continuum are reasons that one might want to work in the discrete.  In 1995, H-P. Nollert and H. Herold developed the wire-frame method which was the first attempt at finding a discrete numerical solution to the embedding problem~\cite{NH:1996}.  Their method used position vectors $\vec{r}(\xi_i(t))$ where $\xi_i$ are the internal parameters of the surface. Given $\vec{r}(\xi_i(t))$, they made a Taylor series expansion around points on the surface allowing them to write distances in Euclidean space in terms of components of the internal metric.  Equating this distance with the actual Euclidean distance gave them a function to minimize where the variables are coordinates in $\mathbb{R}^3$.  Because their optimization function only restricted edge lengths, and in general polyhedrons are not uniquely determined by them, their method had no means of choosing what they called smooth solutions.

In 2008, D. Kane et. al wrote a pseudopolynomial time algorithm to give a numerical realization of Alexandrov's embedding theorem~\cite{Kane:2009}.  Instead of following the proof by Alexandrov, they modeled their algorithm using the proof by A. Bobenko and I. Izmestiev~\cite{Bobenko:2008}.  Given a convex polyhedral metric, this method uses the variation of three dimensional curvatures within the interior of the polyhedron to slowly deform their surface to one that is isometric.  Alexandrov's theorem says given a convex polyhedral metric there exists a unique convex polyhedron in $\mathbb{R}^3$.  Kane et. al showed that their algorithm finds an approximate convex polyhedron where no edge has a dihedral angle greater than $\pi + \epsilon$. The time taken to reach this $\epsilon$-convex polyhedron depends on several intrinsic variables of the metric and user specified tolerances. Their algorithm only works for convex polyhedral metrics. 

In this paper we present the adiabatic isometric mapping (AIM) algorithm which is a numerical approach for embedding polyhedral metrics. Like D. Kane et. al's, our algorithm produces approximately convex polyhedrons given convex polyhedral metrics.  For metrics that are not convex, AIM produces smooth polyhedrons similar to those mention in~\cite{NH:1996}. The AIM algorithm borrows techniques from several of the algorithms mentioned above. The first step of AIM uniformizes the initial polyhedral metric under discrete Ricci flow in an affine time parameter $t$.  
The second step embeds this constant curvature surface near the surface of a sphere.  We use this embedded surface as the starting point for an embedding flow back to the original metric.  This is in a similar vein to M. Jasiulek and M. Korzy\'{n}ski's algorithmm.  To step from one conformally related polyhedron to the next, we use Newton's method to minimize an objective function that depends on the edge lengths of the polyhedral metric and the coordinate distances in $\mathbb{R}^3$.  We use the coordinates of the constant curvature polyhedron as the initial guess for Newton's method to avoid the problem of local minima found in Bondarescu et al.  This puts us close enough to the solution so that we quickly converge to the global minimum.  We then use the newly embedded polyhedron as the initial value data for our next step such that $\Delta t$ is small enough to remain near the global minimum.  This is repeated until we reach the original polyhedral metric.  Taking $\Delta t$ small is not enough to ensure that we will not produce non-smooth solutions as seen by Nollert and Herold.  Because of this, we introduce the convexity routine at each time step which guides our solution toward smooth embeddings. Using our condition on $\Delta t$ and the convexity routine, we introduce the guided adiabatic pull-back which is unique to our algorithm.  We found that AIM is capable of embedding with accuracy in the edge lengths to any desired tolerance above machine precision.   
  
\section{Adiabatic Isometric Mapping (AIM)}\label{sec:aim}
Following the approach of Alexandrov, we begin by finding a continuous family of polyhedral metrics $\rho_t$ with $t\in[0\leq t \leq t_f]$.  A polyhedral metric $\rho$ of a triangulated surface is a list of its triangles by vertices $\{v_i,v_j,v_k\}$ together with an assignment of a length to each edge $\ell_{ij}=\overline{v_iv_j}$. Here, $\rho_0$ is the metric we wish to embed and $\rho_{t_f}$ is a metric with constant Gaussian curvature at each vertex.  Each polyhedral metric in the family $\rho_t$ has $N_0$ vertices, $N_1= 3N_0-6$ edges and $N_2=2N_0-4$ flat triangular faces. The squared edge lengths of $\rho_t$ are given as, 
\begin{equation}
\left\{ \ell^2_{ab}\left(t\right) \right\}
\end{equation} 
where the indices a and b label the vertices of $\rho_t$.  It should be noted that throughout our algorithm the structure of the triangulation of $\rho_t$ always remains the same.  The structure is defined as a list of triangles by vertices.  After we obtain $\rho_t$, we find a polyhedron $P_{t_f}$ which is an isometric embedding of $\rho_{t_f}$ in $\mathbb{R}^3$.  For $P_{t_f}$ to be isometric to $\rho_{t_f}$, the coordinates for the vertices of $P_{t_f}$, 
\begin{equation}
a \longrightarrow \left\{ x_a\left(t_f\right),y_a\left(t_f\right),z_a\left(t_f\right) \right\},
\end{equation}  
must satisfy each of the $N_1$ distance relations,
\begin{equation}
\label{eq:iso}
\ell_{ab}^2\left(t_f\right) = (x_a\left(t_f\right) - x_b\left(t_f\right))^2 + (y_a\left(t_f\right) - y_b\left(t_f\right))^2 + (z_a\left(t_f\right) - z_b\left(t_f\right))^2.
\end{equation}
We specify freely 6 of the $3N_0$ coordinates in order to mod out the translations and rotational degrees of freedom.  This is often done by identifying one of the triangles $\Delta_{abc}$ of $P_{t_f}$ and fixing (1) the three coordinates of vertex $a$, (2) two of the three coordinates of vertex $b$, and (3)  one of the remaining three coordinates of vertex $c$.   
The isometric embedding problem involves a solution to a quadratic system of $3N_0-6$ equations and unknown coordinates.  There are many solutions that satisfy the quadratic equations; as a result, one finds oneself in a ``sea of solutions''  that makes solving this system of equations prohibitively difficult.  Not only does one have to solve the non-linear sparsely coupled system, they also have to find a solution with the desired extrinsic curvature.  Constrained optimization problems of this kind are often costly to solve.  
\begin{figure}[h]
\centering
\centerline{\includegraphics[width=5in]{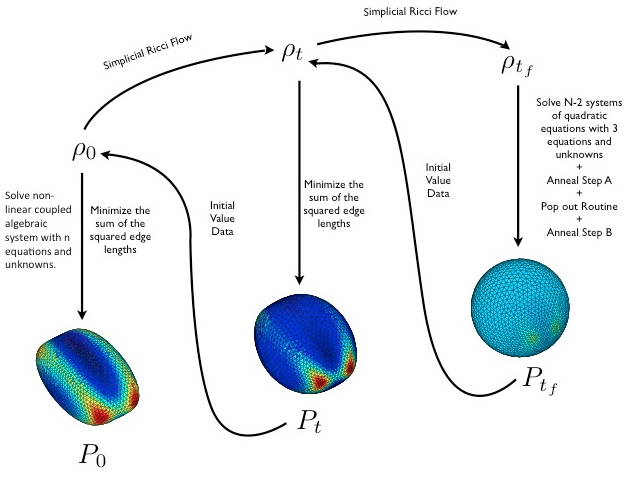}}
\caption{This figure illustrates the three steps of the AIM algorithm.  Sec. 2.1 of our paper,  ``Uniformization via Ricci Flow", begins at $\rho_0$ in the upper left corner and ends at $\rho_{t_f}$ in the upper right corner.  Sec. 2.2, ``Uniform Surface Embedding" begins at $\rho_{t_f}$ and ends at the bottom right of the figure.  Finally Sec. 2.3, ``Guided Adiabatic Pull-Back (GAP)", starts at $P_{t_f}$ in the lower right corner and ends at $P_0$ in the lower left corner of the figure.}
\label{fig:aim}
\end{figure}
To navigate through the ``sea of solutions" without resorting to constrained optimization, AIM uses a three step procedure: (1) uniformization via Ricci flow that provides a dimensional reduction from 3 to 2 dimensions,  (2) uniform surface embedding, and (3) a guided adiabatic pull-back of the coordinates from the uniformed surface $\rho_{t_f}$ to the original surface.  These three steps ameliorate many difficulties in solving the system of equations and provide controllable criteria for obtaining a suitable solution.  We describe these steps in the the next three subections which are illustrated in figure.~\ref{fig:aim}.  From now on when we mention the metric we are referring to a polyhedral metric.

\subsection{Uniformization via Ricci Flow}\label{sec:urf}
We use a discrete Ricci flow to find a conformal relationship between $\rho_0$ and $\rho_{t_f}$.  A conformal factor is assigned to each vertex $a$ of $\rho_0$ and is denoted by the set $\{u_a(t=0)=0\}$.  The relationship between edge lengths of $\rho_0$ and $\rho_t$ is given by,
\begin{equation}
  \ell_{ab}(t) = \ell_{ab}(0) \frac{e^{u_a(t)} +e^{u_b(t)}}{2},
  \label{eq:len}
  \end{equation}
where a and b index the bounding vertices of edge $\ell_{ab}$.

The discrete Ricci flow equations we use  for the conformal factors are given by,   
\begin{equation}
\label{eq:ricci}
\frac{du_a}{dt} = -(k_a - \bar{k}),
\end{equation} 
where $k_a$ is the Gaussian curvature at vertex a and $\bar{k}$ is the target curvature of the surface.  To keep the surface area constant, we choose $\bar{k}$ to be the average Gaussian curvature over the surface, 
\begin{equation}
\overline k := \frac{2\pi \chi}{{\mathcal A}},
\end{equation}
where $\chi$ is the Euler characteristic and $\mathcal A$ is the area of the simplicial surface.  A detailed description of Gaussian curvature can be seen in appendix~\ref{sec:intrincurv}.

It was shown by Chow and Lou that combinatorial Ricci flow and discrete Ricci flow are essentially equivalent and that they exponentially converge to a surface of constant curvature~\cite{CL:discreteRF}.  Therefore given some $\epsilon \ll 1$, it is assured that for sufficiently long times $t_f$ we can Ricci flow the initial conformal factors such that the Gaussian curvature at each vertex is close to $\bar k$,
 \begin{equation}
 || k_a(t_f) -\bar k ||_2 < \epsilon.
 \end{equation} 
 
 Once we have $\{u_a(t_f)\}$, we create a continuous family of conformal factors between $\rho_0$ and $\rho_{t_f}$ using the linear relation,
 \begin{equation} 
 u_a\left(t\right) = u_a\left(t_f\right)\left(\frac{t}{t_f}\right) .
 \label{eq:linlink}
 \end{equation}
 We also considered using an exponential relationship between conformal factors and the set of conformal factors produced at each step during Ricci flow. These alternate paths between conformal factors were shown to be computationally more demanding without any apparent benefit over the linear path.  We discuss this in more detail in Sec.~\ref{sec:GAP}.  
\subsection{Uniform Surface Embedding}\label{sec:unisurf}
The uniform surface Embedding is broken into four steps,
\begin{enumerate}
\item Embedding on a sphere
\item Anneal step A
\item Convexity Routine
\item Anneal step B.
\end{enumerate}
We will now discuss each step in detail and explain their necessity beginning with embedding on a sphere.  

Ideally a constant Gaussian curvature surface will lie on a sphere centered at the origin with radius,
\begin{equation}
\label{eq:avg}
r  = \frac{1}{\sqrt{\overline{k}}}.
\end{equation}
This gives an additional constraint, $r_i^2=x_i^2+y_i^2+z_i^2$, on the coordinates of $\rho_{t_f}$.We are free to specify the initial line segment $\ell_{ab}$ as $\{x_a,y_a,z_a\}=\{\frac{r}{2},\frac{r}{2},\frac{r}{\sqrt{2}}\}$ and
\begin{eqnarray}
\label{eq:xb}
 x_b&=&\frac{r(r^2-\ell_{ab}^2)+\sqrt{\ell_{ab}^2r^2(2r^2-\ell_{ab}^2)}}{2r^2},\\
 \label{eq:yb}
 y_b&=&\frac{r(r^2-\ell_{ab}^2)-\sqrt{\ell_{ab}^2r^2(2r^2-\ell_{ab}^2)}}{2r^2},\\
 \label{eq:zb}
 z_b&=&\frac{r}{\sqrt{2}}.
\end{eqnarray}  
Given this embedded line segment, we embed vertex $c$ by solving,
\begin{eqnarray}
\label{eq:sys1}
\ell_{ac}^2 &=& (x_{a} - x_{c})^2 + (y_{a} - y_{c})^2 + (z_{a} - z_{c})^2,\\
\label{eq:sys2}
\ell_{bc}^2 &=& (x_{b} - x_{c})^2 + (y_{b} - y_{c})^2 + (z_{b} - z_{c})^2,\\
\label{eq:sys3}
r^2 &=& x_{c}^2 + y_{c}^2 + z_{c}^2.
\end{eqnarray}
There are ordinarily two solutions to these equations which correspond to the reflection of vertex $c$ about $\ell_{ab}$.  The translational and rotational degrees of freedom are fixed once we embed the three vertices of $\Delta_{abc}$ on the surface of the sphere.

\begin{figure}  
\centering
\includegraphics[width=4.5in]{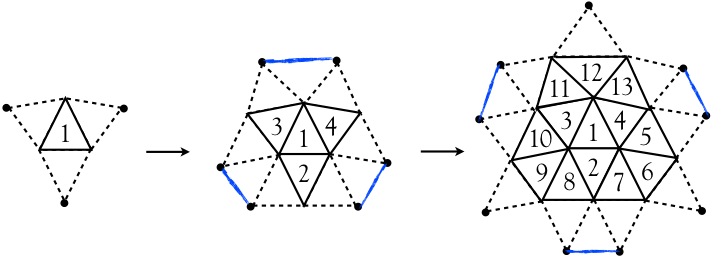}
\caption{Triangles made of solid lines have already been embedded and are used as pivots to solve for the next set of coordinates. Dotted lines represent edges that are used, in conjunction with the radius, to compute the coordinates of the vertex represented by the dots.  The blue paint brush lines, or determined edges, are edges that are determined once the coordinates of its vertices are found.  They are not actually used to compute coordinates. }
\label{fig:tree}
\end{figure}

\begin{figure}
\centering
\includegraphics[width=2in]{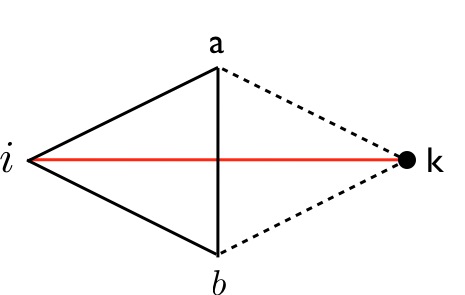}
\caption{Similar to figure~$\ref{fig:tree}$, the dotted lines represent the edges used to compute the coordinates of vertex k.  The vertices of the solid triangle have already been embedded in $\mathbb{R}^3$.  Lastly, the correct solution is selected by maximizing the distance $\bar{ik}$.  This is equivalent to maximizing $\theta_{ab}$ such that it is less than $\pi$.}
\label{fig:lengmax}
 \end{figure}

Starting with this single embedded triangle $\Delta_{abc}$ we can embed each of the three triangles sharing bounding edges.  This procedure is illustrated in figure.~\ref{fig:tree}. In fact given a triangle with a single embedded edge, we embed its opposite vertex by solving (\ref{eq:sys1}) - (\ref{eq:sys3}).  This allows us to generate a growing network of embedded triangles on the surface of the sphere one triangle at a time.  The dimensional reduction afforded to us by the Ricci flow essentially block-diagonalizes the original sparsely-coupled matrix from the quadratic distance equations (\ref{eq:iso}) into many $3\times 3$ matrices.  By looking at shared edges between embedded and non-embedded triangles, we can successively embed each vertex individually until all vertices $a \in \rho_{t_f}$ lie on the sphere.
We choose the solution to these quadratic systems such that the dihedral angle between edge $\ell_{ab}$ is maximal and less than $\pi$.  This solution yields two ``unfolded" triangles embedded on the surface and is illustrated in figure.~\ref{fig:lengmax}.  
 
Here we explain the necessity and logistics behind anneal step A.  The embedding-tree procedure will ordinarily lead to inconsistencies with the isometric embedding equations (\ref{eq:iso}).  The vertices of a constant curvature polyhedron ordinarily will not lie on a sphere of constant radius unless the number of vertices $N_0$ is sufficiently large.  Inconsistencies occur on the ``determined edges'', as illustrated in figure.~\ref{fig:tree}, whose edge lengths do not agree with those in $\rho_{t_f}$.  We correct this disagreement by annealing the coordinates on the sphere to the edge lengths of $\rho_{t_f}$.   Annealing is done using Newton's method to minimize the ${\mathbb L}_2$ norm of (\ref{eq:iso}) at time $t=t_f$,  
\begin{equation}
\sum_{ab}{\left(l_{ab}^2\left(t_f\right) - [x_a\left(t_f\right) - x_b\left(t_f\right)]^2 - [y_a\left(t_f\right) - y_b\left(t_f\right)]^2- [z_a\left(t_f\right) - z_b\left(t_f\right)]^2 \right)^2} = 0,
\label{eq:fmin}
\end{equation}
with the coordinates on the sphere as our initial guess.  The surface after annealing is uniform in the intrinsic sense but extrinsically it may have negative curvature in some places.  Before we move on to fix this issue, we must discuss the isometric variations of polyhedra.

\begin{figure}[ht]
  \centering
    \begin{tabular}{c c}
      \includegraphics[width=.4\linewidth]{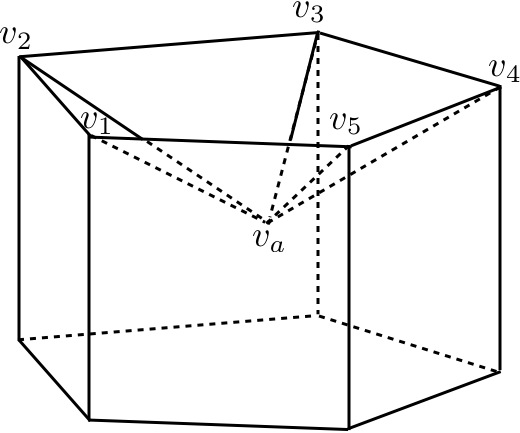} &  \includegraphics[width=.4\linewidth]{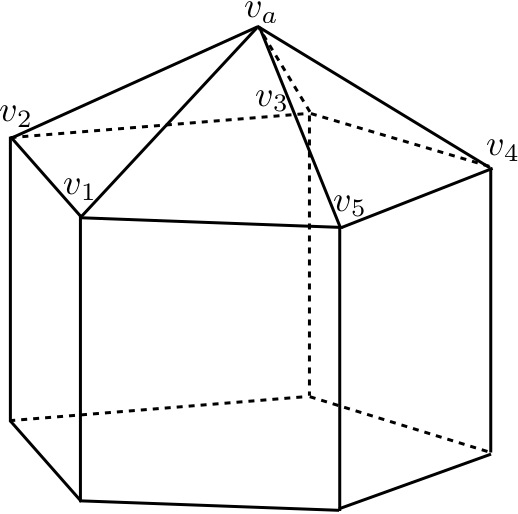} \\
      (a)  &  (b)
    \end{tabular}
\caption[example]
{This is an example of two polyhedron with identical edge lengths and Gaussian curvatures but different extrinsic curvatures.}
 \label{fig:inout} 
 \end{figure}

It is well known that polyhedra are not uniquely determined by their edge lengths since this restriction says nothing about their extrinsic curvature.  For example, figure~\ref{fig:inout} shows two polyhedra with identical edge lengths and Gaussian curvatures but different extrinsic curvatures.
 If all the edges emanating from a vertex have negative extrinsic curvature, that vertex is identified as an inverted vertex.  This is seen in figure~\ref{fig:drumcomp}a.  On average there are six edges emanating from a vertex of an arbitrary triangulated polyhedron. A vertex is identified as partially inverted if more than seventy percent of the edges emanating from it have negative extrinsic curvature.  Identifying these vertices allows us to guide our embedding toward the most convex polyhedron possible during the guided adiabatic pull-back procedure.  The identification of partially inverted vertices was chosen empirically based on experiments with AIM for a variety of surfaces. We expect that this choice is most effective when dealing with triangulations that contain mostly regular triangles.  For example, all of our triangulated surfaces had internal angles strictly between $\frac{\pi}{6}$ and $\frac{\pi}{2}$.
 
An inverted vertex can be ``popped out" such that the edges maintain their lengths but their extrinsic curvatures are no longer negative.  This process is local meaning all other vertices maintain their coordinates and all other edge lengths remain the same. We show this in figure~\ref{fig:drumcomp}b.  On the other hand, a partially inverted vertex can not be removed locally and it is not clear what it means for them to be inverted.  That is why we need the convexity routine during the uniform embedding and GAP procedures to apply a sort of outward pressure that makes the embedding as convex as possible.  The convexity routine is described as follows: first compute the average vector $\vec{v}_{avg}$ of the vertices sharing an edge with the inverted or partially inverted vertex,  second compute the difference vector defined as $\vec{v}_{diff}=\vec{v} - \vec{v_{avg}}$, and third define the new position vector of vertex $v$ by $\vec{v}_{new}=\vec{v}+2\vec{v}_{diff}$.  For an inverted vertex this will get one close to the "popped out" solution were edge-lengths are invariant.  For a partially inverted vertex the edge-lengths are no longer the same, but when annealing, i.e. minimizing in the space of coordinates, our initial guess is now closer to a more convex solution.  In both cases annealing after the convexity routine returns the edge lengths to their original values within some set tolerance, this is the purpose of anneal step B.  The convexity routine may not work for an abritrary triangulation, but should be sufficient for triangulations where the internal angles of all triangles are greater than $\frac{\pi}{6}$ and less than $\frac{2\pi}{3}$.    

\subsection{Guided Adiabatic Pull-Back (GAP)}\label{sec:GAP}

In the final step of the AIM algorithm we pull-back from $P_{t_f}$ to $P_0$ by finding coordinates of $P_t$ for all steps $t_j\in[0,t_f]$.  These coordinates are found by minimizing (\ref{eq:fmin}) at each $t_{f-j}$.  For $t_{f-1}$ we used the uniform embedding found in the previous section as our initial guess.  If $j=2,3$, we use the coordinates at time $t_{f-j+1}$ as the initial guess for embedding surface $\rho_t$.  For $j>3$ we extrapolate the coordinates at $t_{f-j}$ using the previous three coordinate sets.  This extrapolation procedure has two purposes: first it brings our initial guess closer to the global minimum which decreases the convergence time for Newton's method, second it allows us to take larger steps from $t_{f-j}$ to $t_{f-j-1}$ while keeping us within an open ball of the global minimum.  This better initial guess further decreases the run time of GAP.

There are several paths that relate conformal factors $\{u_a\left(t_f\right)\}$ to $\{u_a\left(0\right)\}$.  We chose the linear path given by (\ref{eq:linlink}).  This path is optimal because the rate of change of Gaussian curvature is constant at each time step $t_{f-j}$ to $t_{f-j-1}$, which allows one's step size $\Delta t = \frac{t_f }{Tsteps}$ to be constant throughout GAP.  Here $Tsteps$ is the number of steps taken.  If the path was not linear, we would need adaptive time stepping to account for the variable change in Gaussian curvature.  For example, one of the alternative paths that we considered was given by the conformal factors produce at each step of Ricci flow.  Since Ricci flow exponentially uniformizes curvature, there is a greater change in curvature near $t=0$.  Therefore it is necessary to decrease $\Delta t$ to maintain adiabaticity near this region of rapid change.  We also considered an exponential path given by the time interval, $t_j = \left(\frac{e^{-\alpha Tsteps} - e^{-\alpha j}}{e^{-\alpha Tsteps} - 1}\right)t_f$ where $\alpha$ is the parameter that controls the rate of convergence.  The exponential paths suffers the same problem in adaptive stepping as the Ricci flow path.  Neither of these paths yielded improved results over the linear path and both are computationally more demanding.

\begin{figure}[h]
\centering
\includegraphics[width=4in]{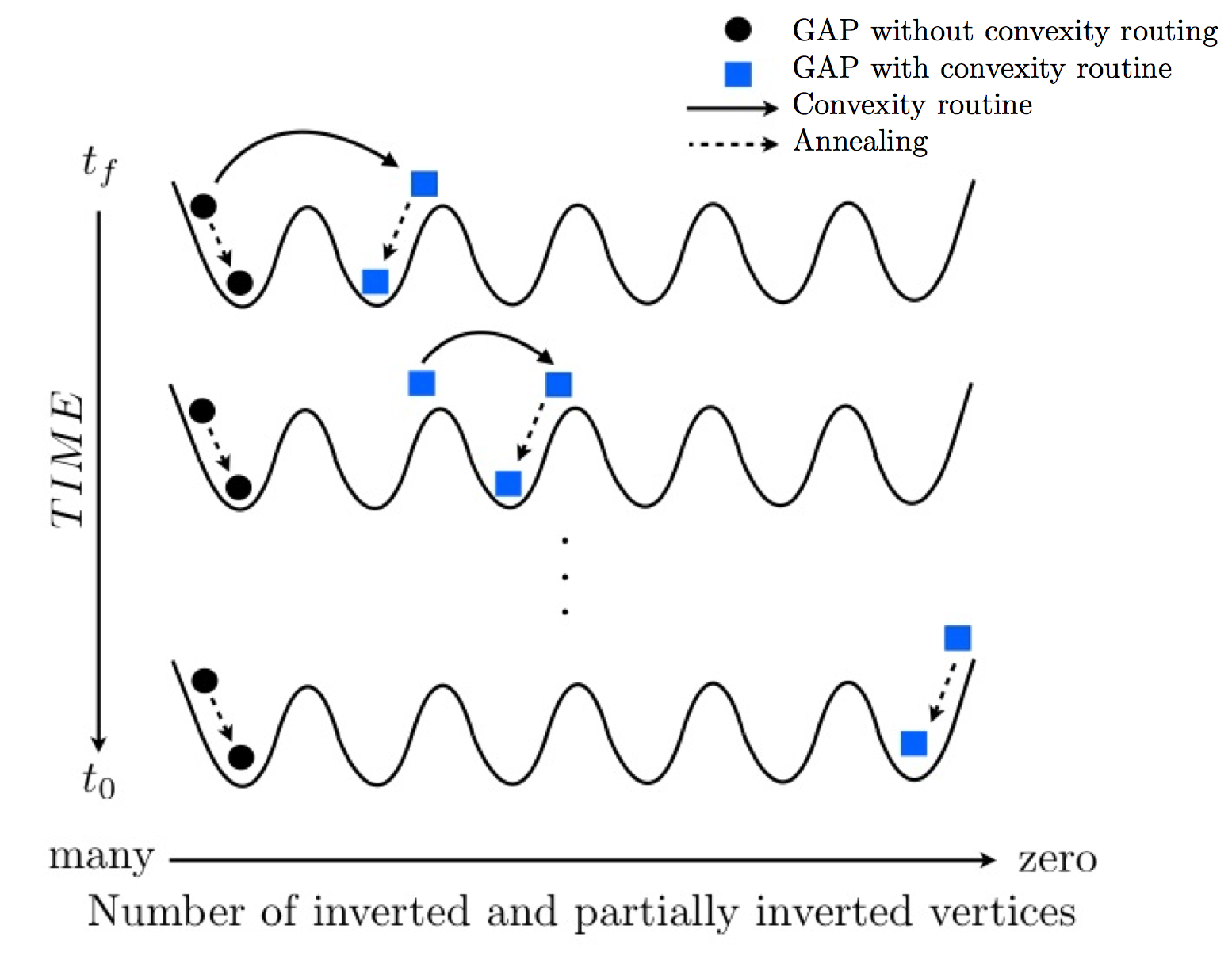}
\caption{This is a overly simplified visual representation of minimization in coordinate space.  Each application of the convexity routine perturbs us away from undesirable global minima.  After many time steps we eventually reach a global minimum without any inverted or partially inverted vertices.}
\label{fig:cspace}
\end{figure}

As we mentioned in the previous section, minimization techniques can not determine extrinsic curvature.  Even after applying annealings and the convexity routine to $P_{t_f}$, it is still possible to begin with an initial surface at $t_f$ that has many clustered and isolated partial inverted vertices.  If nothing was done GAP would produce an undesired crumpled surface surface.  We apply the convexity routine at each step to avoid caved in crumpled polyhedra.  Within the space of coordinates, this perturbs our initial guess such that we begin near solutions without partially inverted vertices.  Although this does not immediately remove the problem vertices, it gradually reduces them at each step until they are eliminated.  We illustrate this gradual evolution away from crumpled embeddings in figure~\ref{fig:cspace} keeping in mind that (\ref{eq:fmin}) is nonlinear thus having several global minima whose residuals are zero.  
Using all of these elements, we defined our evolution as adiabatic if $\Delta t$ is small enough such that we transition to and remain near a global minimum with no inverted or partially inverted vertices.  This gradual nudging during GAP and its ability to move toward a more convex polyhedra is the main result of this manuscript.

\section{Numerical Tests}\label{sec:num}
We present three numerical examples of the AIM algorithm. The first two example meshes were made using the distmesh program \cite{tri:distmesh}.   As input, distmesh is given $z\left(x,y\right)$ to produce coordinates and a list of triangles, by vertices, for surfaces in $\mathbb{R}^3$.  We use these coordinates to get edge lengths by calculating the Euclidean distance between vertices.  We then use the extracted edge lengths together with the list of triangles as the original polyhedral metric for which we apply AIM.  Once AIM is complete, we check how well AIM reproduces intrinsic curvature by comparing edge lengths of the distmesh surface to those produced AIM.  We compare extrinsic curvature by looking at the convergence of integrated mean curvature produced by AIM to the continuum value.  

Our third example is a surface just outside the ergosphere of a Kerr black hole that is right below maximal rotation.  We could not use distmesh to triangulate this surface since its embedding equations are quasi-analytic, which makes them incompatible with distmesh's input format.  Because of this, we triangulated the ergosphere ourselves. 

\subsection{Distmesh surfaces}\label{sec:distmeshnum}
The first two surfaces we chose as our test cases are given by,
\begin{gather}
\label{eq:ellipse}
\frac{x^2}{a^2}+\frac{y^2}{b^2}+\frac{z^2}{c^2}=1, \left(Ellipsoid\right)\\ 
\centering
\label{eq:drum}
\frac{x^2}{d^2} + \frac{y^2}{e^2}+ \frac{z^4}{f^4} = 1 \left(Drum\right)
\end{gather}
where $a=3$, $b=2$, $c=1$, $d=2$, $e=1$ and $f=\sqrt{1.5}$.  Both surfaces have strictly positive point wise Gaussian and mean curvatures.  We will refer to the surfaces described by (\ref{eq:ellipse}) and (\ref{eq:drum}) as the ellipsoid and drum, respectively. 

The following work was programmed using matlab.  This preliminary application of AIM does not attempt to optimize our algorithm.  The goal is to verify our approach and test how accurately it preserves distances and curvature quantities.  Run times can be significantly improved by parallelizing the code and using preconditioning for our optimization routines. 
 \begin{figure}[hp]
\includegraphics[width=5in]{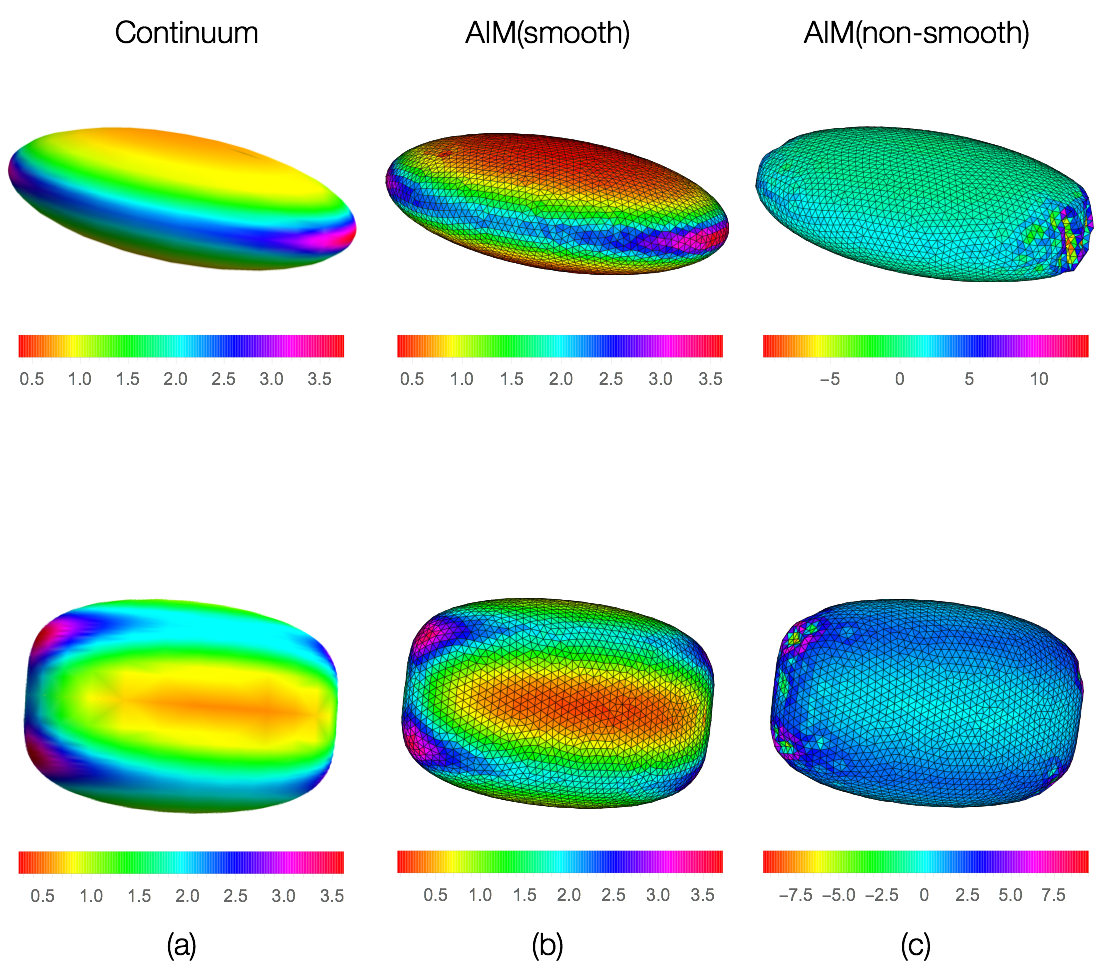} 

\caption
{On the left we have the embedding using the continuum equations.  In the center we have the smooth embeddings produced by AIM.  On the right we have a non-smooth embeddings which are produced when we do not pull-back adiabatically and we do not use the convexity routine.  The color coding indicates the mean curvature associated to each triangle.  The ellipsoid has a resolution of 3226 vertices while the drum's resolution is 3246.}
 \label{fig:drumcomp} 
 \end{figure}
When using AIM, one is allowed to choose the accuracy of their embedding by manipulating the tolerances during the annealing and GAP procedures.  Tolerances are set using the value of the residuals and the magnitude of the steps in coordinate space for each iteration of the quasi-Newton method.  These tolerances are set to $10^{-6}$ and $10^{-8}$, respectively.  This means the Newton's method will stop if the edge lengths are within $10^{-6}$ accuracy or if the distance between points in coordinate space from one iteration to another is on the order of $10^{-8}$.  Given these settings, we determine the number of steps necessary for maintaining adiabaticity when performing GAP. We also present our analysis on how each portion of the code scales with increased vertices, as well as the scaling for the number of time steps necessary for adiabaticity.  For each surface tested, we recovered an average difference in the edge length between the original metric and AIM within our prescribed tolerance.  The standard deviation in the difference between edge lengths are also on the order of our prescribed tolerance. This result was tested with tolerances in accuracy set to $10^{-6}$ and $10^{-9}$.  Our reported scaling analysis corresponds to a tolerance of $10^{-6}$. 

\begin{figure}[ht]
\begin{center}
\includegraphics[trim=15mm 60mm 15mm 60mm,clip,width=4in]{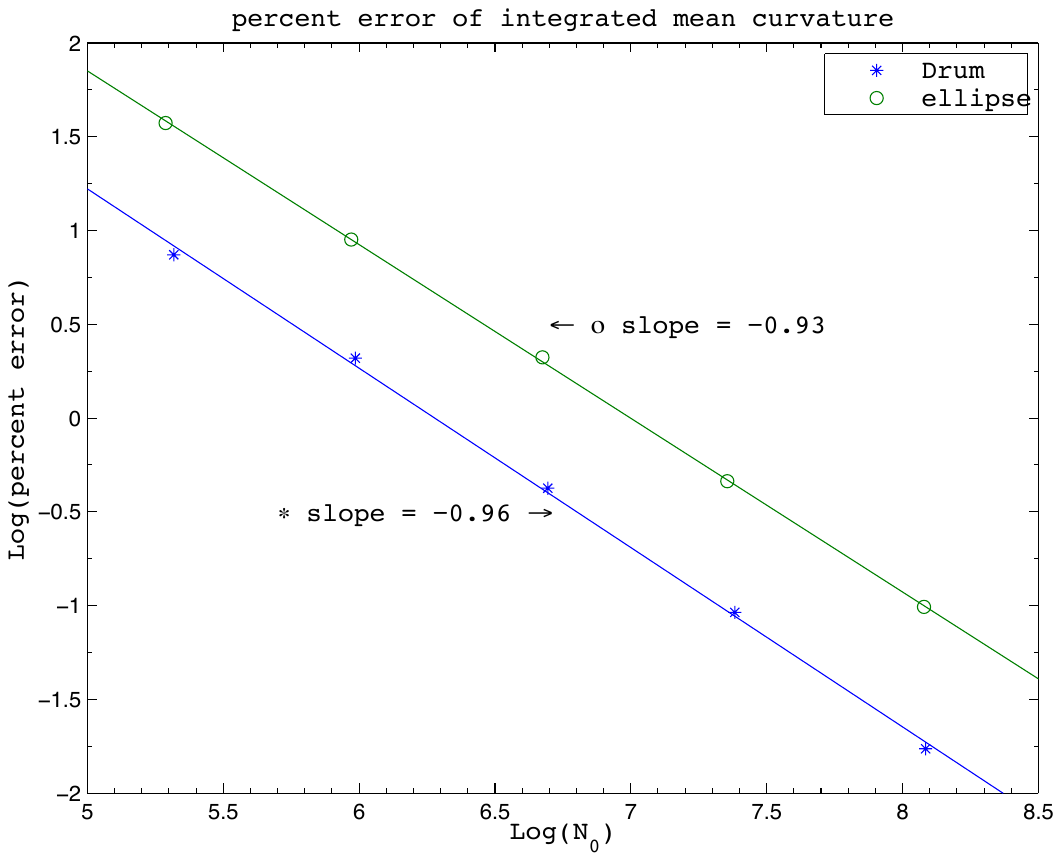}
\caption{Convergence of integrated mean curvature as a function of resolution}
\label{fig:conv}
\end{center}
\end{figure}

We test whether or not the extrinsic curvature is recovered by first looking at a qualitative comparison between continuum and AIM embeddings as seen in figure~\ref{fig:drumcomp}.  These surfaces are color coded as a function of mean curvature.  The mean curvature of the continuum surfaces were computed using the continuum equations while the triangulated mean curvatures were computed using discrete methods~\cite{DiscMC}.  Figure~\ref{fig:conv} gives a quantitative comparison, and convergence properties, of the mean curvature between the continuum and AIM by plotting the percent error in integrated mean curvature as a function of resolution.  According to our results, the convergence of the integrated mean curvature to the continuum is of first order.     

\begin{figure}[ht]
\begin{center}
\includegraphics[trim=15mm 60mm 15mm 60mm,clip,width=5.1in]{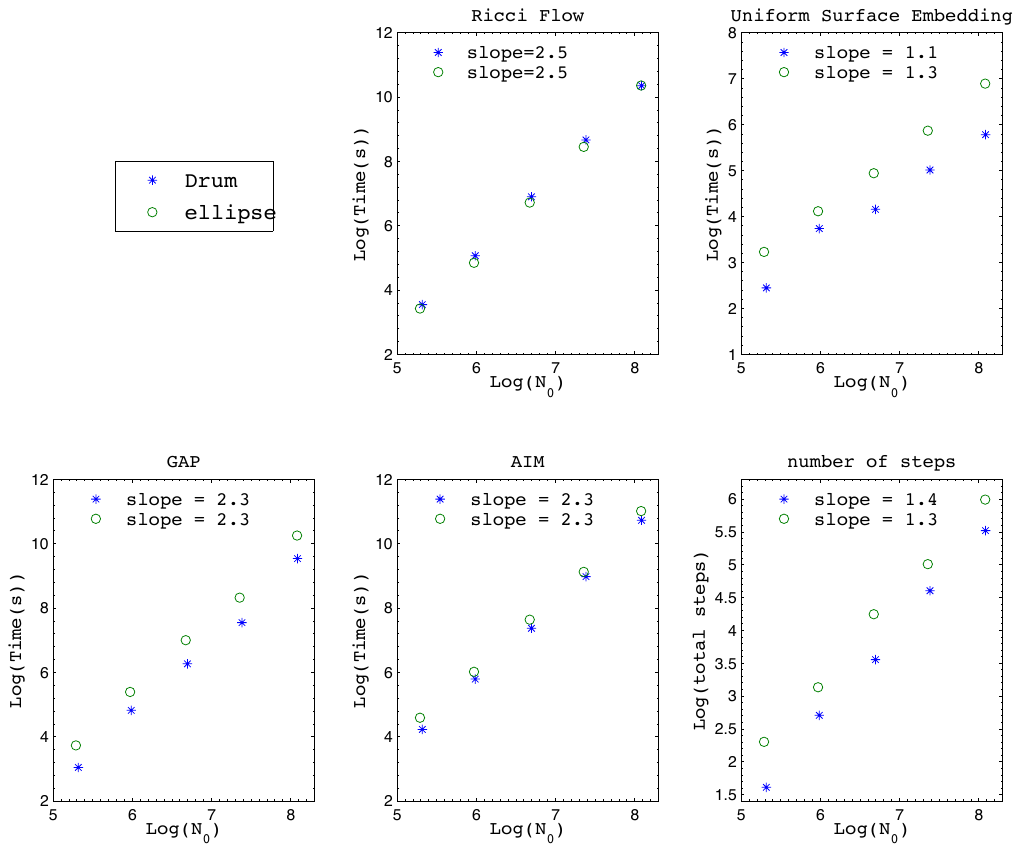}
\caption{These are log-log plots of each section of AIM and the algorithm as a whole.  The slopes indicate scaling behavior of our algorithm as a function of $N_0$.  Our overall results suggest that AIM scales sub cubically with time. Time is in units of seconds.}
\label{fig:plots}
\end{center}
 \end{figure}

We made log-log plots of run time verses number of vertices for each of the three sub routines and the AIM algorithm as a whole to understand the scaling behavior of our code. The same plots were made for analyzing the number of steps necessary for adaibaticity as a function of resolution.  The plots are given in figure~\ref{fig:plots}.  It is observed that the highest order contribution to the scaling of AIM comes from the Ricci flow procedure.  Our overall results suggest that AIM scales sub cubically. 

\subsection{Modified Ergosphere of Kerr spacetime}\label{sec:ergo}
\begin{figure}[h]
  \centering
  \includegraphics[width=5.5in]{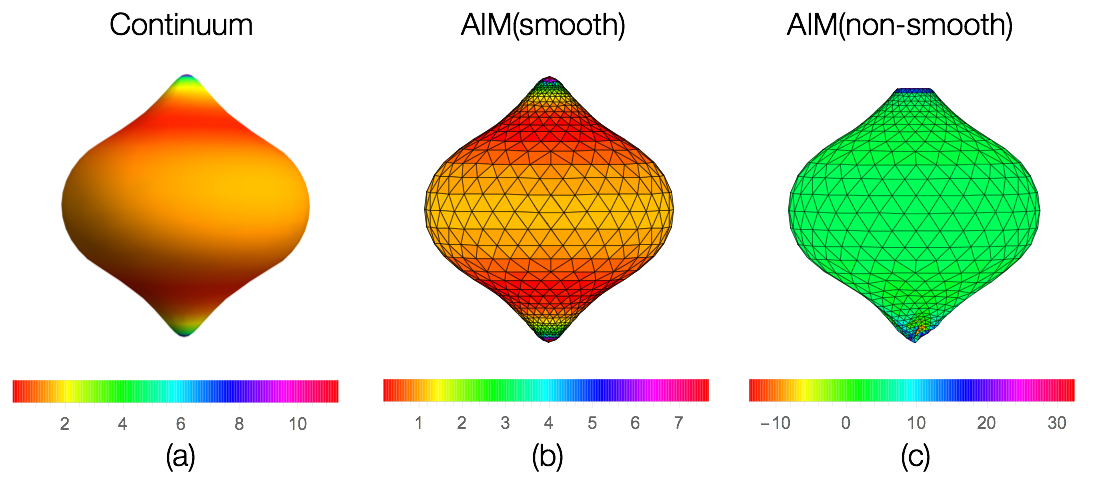} 
\caption
{On the left we have the original embedding made using the continuum embedding equations.  The middle is the smooth embedding given by AIM.  On the right we have the non-smooth embedding that is produce when one does not take small enough time steps and does not use the convexity routine.}
 \label{fig:ergo} 
 \end{figure}

In 1973 Larry Smarr analytically embedded an axisymmetric 2-surfaces of a rotating black hole geometry~\cite{Smarr:1973}.  Following this we use the Boyer-Lindquist coordinates for the metric representation of the rotating black hole spacetime,

\begin{equation} 
 ds^2=g_{tt}dt^2 + 2g_{t\phi}dtd\phi + g_{rr}dr^2 + g_{\theta \theta}d\theta^2 + g_{\phi \phi}d\phi^2,
 \label{eq:ergomet}
 \end{equation}
 where
\begin{eqnarray}
\label{eq:gtt}
g_{tt}&=&-\left(1-\frac{2Mr+Q^2}{\Sigma}\right) ,\\
 \label{eq:gtp}
g_{t\phi}&=&-\frac{2Mr-Q^2}{\Sigma}a \sin^2{\theta} ,\\
 \label{eq:grr}
 g_{rr}&=&\frac{\Sigma}{\Delta},\\
 \label{eq:gthth}
 g_{\theta \theta}&=&\Sigma,\\
 \label{eq:gpp}
 g_{\phi \phi}&=&\left(r^2 + a^2 + \frac{\left(2Mr-Q^2\right)a^2 \sin^2{\theta}}{\Sigma}\right)\sin^2{\theta}.
\end{eqnarray}  
Here we used the usual definitions where 
\begin{eqnarray}
\label{eq:sigma}
\Sigma&:=&r^2 + a^2 \cos^2{\theta},\\
\label{eq:delta}
\Delta&:=&r^2 -2Mr +a^2 +Q^2.
\end{eqnarray}
 For our third example of the AIM algorithm we embedded a distorted ergosphere of a nearly maximally rotating black hole geometry with zero charge.  This surface is defined on a $t=const$ hypersurface thus making $dt=0$.  We have $a/M=0.975$ with $a=0.975$ and $M=1$ where $a/M=1$ gives a maximally rotating surface.  Given that the surface is axial symmetric, we can write $r=R\left(\theta\right)$ which implies that $dr=R_{,\theta}d\theta$.  Inserting this into (\ref{eq:ergomet}) we get the two metric,
 \begin{equation}
 \label{eq:twosurf}
 ds^2=\underbrace{\Sigma\left(1+\frac{R_{,\theta}^2}{\Delta}\right)}_{h_{\theta\theta}}d\theta^2 + \underbrace{\left(R\left(\theta\right)^2 +a^2 +\frac{2a^2MR\left(\theta\right)\sin^2{\theta}}{\Sigma}\right)\sin^2{\theta}}_{h_{\phi\phi}} d\phi^2.
 \end{equation}
 The ergosphere is defined by the radius in which $g_{tt}=0$ in (\ref{eq:ergomet}).  At this point the surface is elongated at the poles which makes it an extreme surface to embed.  To make the poles more smooth we distort the surface by adding a small parameter $\epsilon$ that gives us a surface slightly outside the ergosphere. The radius as a function of $\theta$ is given as,
 \begin{equation}
 \label{eq:rtheta}
 r=R\left(\theta\right)=\underbrace{M\left(1+\sqrt{1-\left(\frac{a}{M}\right)^2 \cos^2\theta}\right)}_{r_{ergo}} + \epsilon M.
 \end{equation}
 Let the isometric embedding functions be defined as,
  \begin{eqnarray}
  \label{eq:x}
 x\left(\theta,\phi\right)&=&\rho\left(\theta\right)\cos\phi,\\
 \label{eq:y}
 y\left(\theta,\phi\right)&=&\rho\left(\theta\right)\sin\phi,\\
 \label{eq:z}
 z\left(\theta\right)&=&f\left(\theta\right).
 \end{eqnarray}
 Equating the line elements between spaces gives the relation,
 \begin{equation}
 \label{eq:linseg}
 ds^2=dx^2 + dy^2 + dz^2 = \left(\rho_{,\theta}^2 + f_{,\theta}^2\right)d\theta^2 + \rho^2d\phi^2.
 \end{equation}
 Using (\ref{eq:twosurf}) and (\ref{eq:linseg}) to solve for $\rho$ and $f$ we have,
 \begin{eqnarray}
 \label{eq:rho}
 \rho\left(\theta\right) &=& \sqrt{h_{\phi \phi}\left(\theta\right)} ,\\
 \label{eq:f}
f\left(\theta\right)&=&\int_0^\theta \sqrt{h_{\theta \theta} - \frac{h_{\phi \phi,\theta}^2}{4h_{\phi \phi}}}.
 \end{eqnarray}
 These equations give us our continuum embedding in $\mathbb{R}^3$. The function $f\left(\theta\right)$ is a real valued function on the domain $\theta \in [0,\pi]$.  We construct the polyhedral metric for the ergosphere by first building a list of $\theta$'s and $\phi$'s for the vertices.  Next we construct a list of triangles by vertices and assign edge lengths by isometrically embedding the vertices of edges and computing the Euclidean distance.  After assigning edge lengths, we discard the coordinates and apply AIM to our new polyhedral metric.  Figure~\ref{fig:ergo} illustrates the continuum embedding, the smooth embedding given by AIM and the non-smooth embedding.  The continuum embedding is a parametric plot, in Euclidean space, using (\ref{eq:x})-(\ref{eq:z}) with $\rho$ and $f$ defined by (\ref{eq:rho}) and (\ref{eq:f}), respectively.  
 The percent error of the integrated mean curvature between the continuum and AIM embeddings is $0.62$ percent, which is comparable to the percent error for the drum at the same resolution.  Although we did not analyze the convergence of integrated mean curvature for this surface, we expect its convergence to be similar to the ellipsoid and drum surfaces. Our main obstacle in completing this convergence analysis is producing consistent high quality triangulations for increasing resolution.  The edge lengths are recovered up to the set accuracy of $10^{-6}$.  	
	
\section{Conclusion}\label{sec:conc}
The AIM algorithm was developed using several techniques from previous isometric embedding algorithms.  Our tests of AIM included non-axial symmetric surfaces with strictly positive Gaussian curvature.  We also tested a surface just above the ergosphere of a Kerr black hole, which is axial-symmetric with regions of negative Gaussian curvature.  

The main innovation of the AIM algorithm comes from the guided adiabatic pull-back.  Although we used triangulated surfaces, similar to the wire mesh of Nollert and Herold, GAP allows us to distinguish between the global minima of smooth and non-smooth polyhedra.  GAP also prevents the problem of getting trapped in local minima, as seen by Bondarescu, Alcubierre and Seidel, during Newton's method .  For the surfaces tested, we were capable of reaching residuals on the order of our prescribed tolerance.  AIM also uses an embedding flow similar to Jasiulek and Korzy\'{n}ski's, but is not restricted to strictly positive scalar curvature surfaces.   

Although AIM does not necessarily produce the convex polyhedron given a convex polyhedral metric, we do produce embeddings that recover much of the extrinsic behavior see in the continuum embedding.  Although there are many approximate polyhedral representations of the convex Alexandrov polyhedron that our algorithm can produce, we observe that they do not vary significantly between each other.  We find that this is especially true at higher resolutions.  Our future plan is to provide an upper bound on the deviation in extrinsic curvature between these essentially equivalent embeddings. 

It was not our goal or intention to optimize the AIM algorithm.  Its run time and scaling may be significantly improved by using preconditioning for the optimization routines and parallelization. We would also like to develop an algorithm that produces well posed triangulation, similar to those made by distmesh, given any compact two metric.  Since our original interest in isometric embeddings stemmed from computing quasi-local energy in general relativity, we plan to use AIM for this purpose.  It is also noted that given a suitable triangulation algorithm, AIM should be readily applicable to two surfaces identified within numerical relativity codes that use finite differencing grids.  One can interpolate to obtain the proper distances of the edges for the triangulation using the grid's metric and coordinates.  This polyhedral metric can then be embedded into $\mathbb{R}^3$ using AIM.

\section*{Acknowledgements}
This work was supported from Air Force Research Laboratory (AFRL/RITA) Grant \# FA8750-11-2-0089 and \# FA5750-15-2-0047. WAM, and SR acknowledge support from Air Force Office of Scientific Research through the American Society for Engineering EducationÕs 2012 Summer Faculty Fellowship Program, and from AFRL/RITA and the Griffiss InstituteÕs 2013 Visiting Faculty Research Program. Any opinions, findings and conclusions or recommendations expressed in this material are those of the author(s) and do not necessarily reflect the views of the AFRL.

\appendix
\section{Intrinsic Curvature Calculations}\label{sec:intrincurv}
In Sec. \ref{sec:urf} we used simplicial Ricci flow to obtain a triangular mesh with constant Gaussian curvature at each vertex.  This procedure relied on curvature values defined on the surface of our lattice.  The purpose of this appendix is to inform the reader about the nature of curvature on simplicial geometries and provide the exact equation used to compute Gaussian curvature.  

It was established by Alexandrov, and utilized by Tullio Regge~\cite{regge:GR}, that the intrinsic curvature of a simplicial geometry, of arbitrary dimensions, is concentrated at co-dimension 2 simplices called hinges.  The curvature at these hinges is a conic singularity whose value goes to infinity as the area of the loop of parallel transport enclosing the hinge goes to zero.  Later it was determined by W. Miller et. al.~\cite{miller:hilbert,srf} that given a Daulanay lattice $\mathcal{S}$ with a circumcentric Voronoi dual lattice $\mathcal{S}^*$, it was possible to define a unique path of parallel transport such that the sectional and Gaussian curvatures are defined.  The area $h^*$ of this loop is the Voronoi area perpendicular to a hinge in the Daulanay lattice.  The Voronoi area $h^*$ is constructed using dual edges $\lambda \in \mathcal{S}^*$ whose vertices are defined as the circumcenters of d-simplices.  If one were to parallel transport a vector around this loop, they would find their vector rotated by the deficit angle at the hinge.  The deficit angle is defined as $\epsilon_h = 2\pi - \sum_i^n{ \eta_i}$, where  the sum is over the interior angles of d-simplicies that share the hinge h.  Using the deficit angle and the dual area associated with h, the Gaussian curvature is defined as,

\begin{equation}
\label{eq:gauss}
k_h = \frac{\epsilon_h}{|h^*|}
\end{equation}   
Everything mentioned in this appendix is applicable for arbitrary dimensions.   Since we are embedding an $\mathcal{S}^2$ surface, we are only concerned with two dimensions.  This means the curvature is concentrated on vertices.

\bibliography{aim2} {}
\bibliographystyle{unsrt}
 
\end{document}